\documentclass[twocolumn,nofootinbib]{revtex4-2}
\usepackage[dvipsnames]{xcolor}
\usepackage{tikz}
\usetikzlibrary{shapes.misc}

\usepackage{amsmath,amssymb,graphicx}
\usepackage{soul,xcolor}
\usepackage{url}
\usepackage[colorlinks,urlcolor=blue]{hyperref}

\begin{document}

\title{Diffusion wake: a distinctive consequence of the Mach-cone wake induced by supersonic jets in high-energy heavy-ion collisions}

\author{Zhong Yang$^{1,2}$}
\email[]{zhong.yang@vanderbilt.edu}
\author{Xin-Nian Wang$^1$}
\email[]{xnwang@ccnu.edu.cn}
\affiliation{$^1$Key Laboratory of Quark and Lepton Physics (MOE) \& Institute of Particle Physics, Central China Normal University, Wuhan 430079, China}
\affiliation{$^2$Department of Physics and Astronomy, Vanderbilt University, Nashville, TN}


\begin{abstract}
In this Research Perspective, we briefly review the diffusion wake, a distinctive consequence  of the Mach-cone wake induced by the supersonic jets in ultra-relativistic heavy-ion collisions. The diffusion wake depletes soft hadrons in the direction opposite to the propagating jet. According to coupled transport and hydrodynamic simulations, a valley in the 2-dimensional  jet-hadron correlation in azimuthal angle and rapidity arises on the top of the multiple parton interaction ridge as an unambiguous signal of the diffusion wake induced by $\gamma$-jets in heavy-ion collisions. In dijet events with a finite rapidity gap, the rapidity asymmetry of the jet-hadron correlation has been shown to be a robust signal of the diffusion wake. The same rapidity asymmetry can also be applied to $\gamma$-jet events  and both are background free. Experimental measurements of these signals can provide valuable insights into the properties of the quark-gluon plasma (QGP) formed in high-energy heavy-ion collisions.

\end{abstract}

\maketitle

Jets are collimated sprays of particles emerging from energetic quarks and gluons in high-energy collisions. Jet-quenching due to jet-medium interaction and parton energy loss \cite{Wang:1992qdg,Cao:2020wlm} is an important phenomenon that can be used to probe properties of the quark-gluon plasma (QGP) formed in high-energy heavy-ion collisions. Jet-medium interaction also gives rise to a medium response induced by the supersonic jet whose velocity (velocity of light) is much faster than the sound velocity inside QGP. The jet-induced medium response consists of a Mach-cone-like wake~\cite{Casalderrey-Solana:2004fdk} in the  jet direction followed by the diffusion wake in the opposite direction of the jet,  leading to perturbations in the soft hadron spectra from the bulk medium and modification of various jet substructures such as the jet shape~\cite{Yang:2022nei}, jet fragmentation functions~\cite{Chen:2020tbl, Casalderrey-Solana:2016jvj}, and jet energy-energy correlators~\cite{Yang:2023dwc, Xing:2024yrb}, among others. Since these perturbations evolve together with the QGP medium, experimental measurements can provide valuable insights into the intrinsic and transport properties of the QGP~\cite{Neufeld:2008dx}.   

Microscopically,  jet-medium interaction transports a medium particle into a recoil particle along the jet direction and leave behind a particle "hole" in the original phase-space of the medium particle. Further propagation of the recoil particles and their interaction with the medium lead to the Mach-cone-like wake while the evolution of the perturbation due to the particle-hole
offset by some of the recoil particles leads to the diffusion wake \footnote{Note that the terms ``Mach-cone-like wake" and ``diffusion wake" we use here refer to the general medium-response along the jet direction that carries energy and momentum deposited by the jet and the depletion of medium behind the jet, respectively.  They are not the same as the ``sound mode'' and ``diffusion mode'' in the linear hydrodynamics response study of the jet-induced wakes~\cite{Casalderrey-Solana:2020rsj}.}. In the direction of a propagating jet, the  Mach-cone-like wake can lead to enhancement of correlated soft-hadrons. However, medium-induced gluon radiation also enhances soft hadrons along the jet direction \cite{Cao:2020wlm}, making it difficult to isolate the contribution from the Mach-cone wake in the final state. As a companion of the Mach-cone-like wake, the diffusion wake, on the other hand, depletes soft hadrons in the opposite direction of the propagating jet \cite{Chen:2017zte, Yang:2021qtl}  which is a unique consequence of Mach-cone-like wake induced by the supersonic jets.   If identified and measured in high-energy heavy-ion collisions,  signals of the diffusion wake can help us confirm the existence of jet-induced Mach-cone-like wake and study properties of the QGP medium.

Early investigations have shown that the diffusion wake leads to a small amount of suppression of jet-hadron azimuthal angle correlations in the $\gamma$ direction of $\gamma$-jet events in Au+Au collisions compared to p+p at the RHIC energy \cite{Chen:2017zte}. However, hadrons from initial multiple parton interactions (MPI) with a uniform azimuthal distribution, which are negligible at the RHIC energy, can overwhelm the signal of the diffusion wake in Pb+Pb collisions at the LHC~\cite{Yang:2022nei}. Separating and subtracting the MPI contribution is then critical.

Since jets are three-dimensional structures, it is natural to extend the jet-hadron correlation in azimuthal angle to a two-dimensional distribution in both azimuthal angle and rapidity to look for signals of the diffusion wake. Such a 2D jet-hadron correlation has been investigated in $\gamma$-jet events \cite{Yang:2022nei} in which the photon does not interact with the QGP and therefore does not interfere with the signal of the diffusion wake in the photon direction or the opposite direction of the jet. In this study, $\gamma$-jet production in p+p is simulated with the PYTHIA8~\cite{Sjostrand:2014zea} model and in central 0-10\% Pb+Pb collisions within the CoLBT-hydro model~\cite{Chen:2017zte}, respectively, at $\sqrt{s_{\rm NN}}=5.02$ TeV. 
The Linear Boltzmann Transport (LBT) model \cite{Li:2010ts,He:2015pra,Luo:2023nsi} is based on Boltzmann equations for the evolution of both jet shower partons and the recoil partons. The collision kernels contain the complete set of leading order (LO) elastic $2\to 2$ scattering processes while inelastic processes  $2\rightarrow 2+n$ with multiple gluon radiation are described by the high-twist formula \cite{Guo:2000nz,Wang:2001ifa} for induced gluon emission that contains Landau-Pomeranchuck-Migdal interference. The CoLBT-hydro model couples LBT model and event-by-event (3+1)D CCNU-LBNL viscous hydrodynamic (CLVisc) model~\cite{Pang:2018zzo} for a concurrent evolution of parton shower and QGP medium. The coupling is through a source term in the hydrodynamic equations made of soft partons including particle holes or ``negative'' partons from LBT while energetic partons propagate through the hydrodynamic medium which are updated in real time. Jets are reconstructed from final hadrons using the FASTJET algorithm~\cite{Cacciari:2011ma}.

\begin{widetext}
\begin{figure}[h!]
\centering
\includegraphics[width=6.5in]{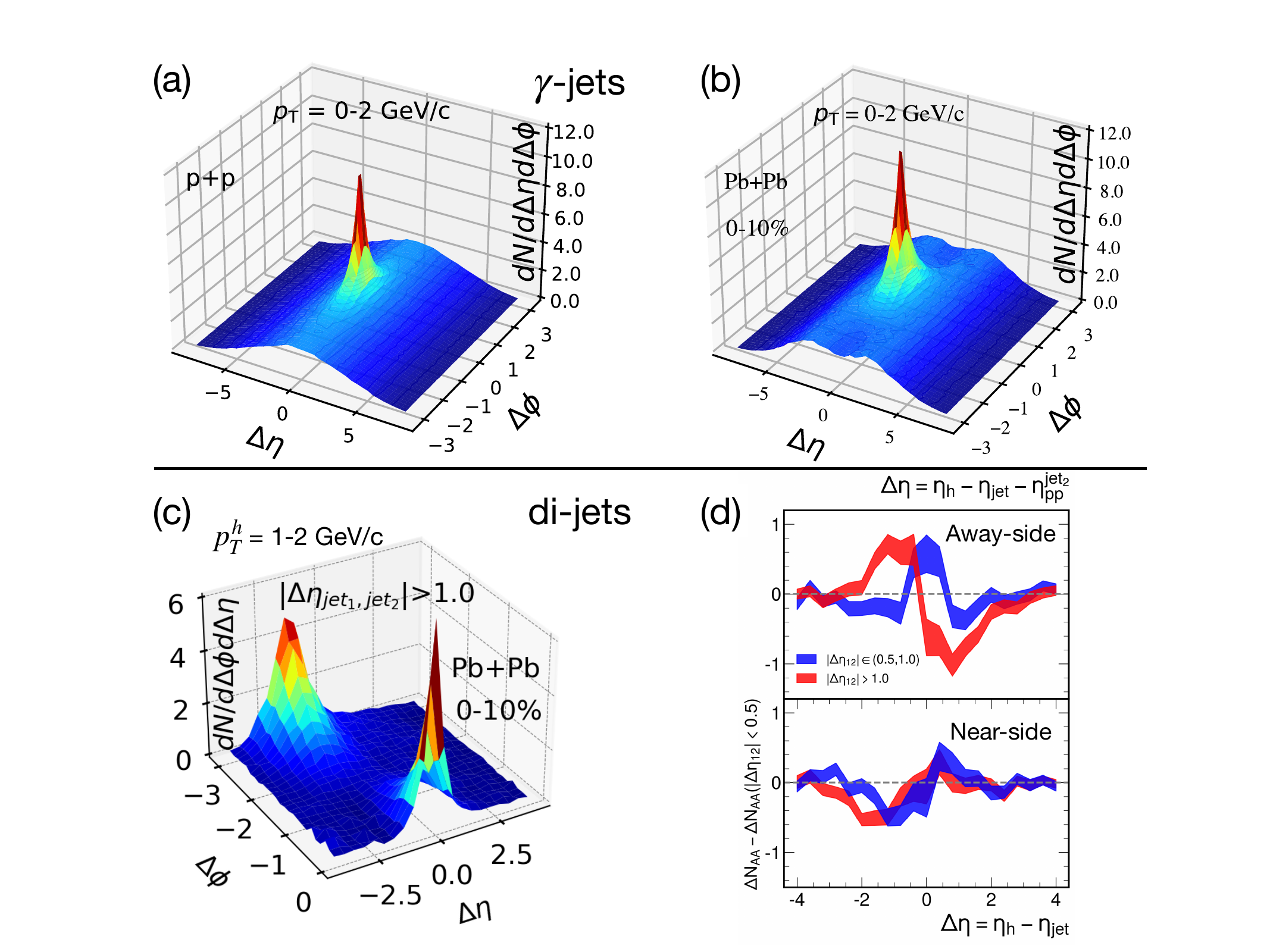}
\label{fig1}
\end{figure}
\centering
\begin{minipage}{0.9\textwidth}
Fig.~\ref{fig1}(a) Jet-hadron correlation in azimuthal angle and rapidity in $\gamma$-jets in p+p, (b) 0-10$\%$ Pb+Pb and (c) in dijets in 0-10$\%$ Pb+Pb collisions at $\sqrt{s}=5.02$ TeV. (d) Rapidity asymmetry of near-side (lower) and away-side (upper) jet-hadron correlation in dijets in 0-10$\%$ Pb+Pb collisions.
\end{minipage}
\end{widetext}

The jet and soft hadron correlations in azimuthal angle and rapidity from $\gamma$-jets in p+p collisions (a) and 0–10$\%$ Pb+Pb collisions (b) at $\sqrt{s}=5.02$ TeV from Ref.~\cite{Yang:2022nei} are shown in the upper panels of Fig.~\ref{fig1}. In p+p collisions, a peak in the jet direction originates from jet hadronization on top of a ridge-like structure along the azimuthal angle direction. This ridge can be attributed to hadrons from MPI which are uniform in azimuthal. In central Pb+Pb collisions, the jet peak is enhanced compared to p+p due to jet-induced medium response and medium-induced gluon radiation. In the direction of the trigger photon or the opposite side of the jet peak, a valley on top of the MPI ridge is formed, resulting from the depletion of soft hadrons by the jet-induced diffusion wake. This valley on top of the MPI ridge provides a clear and unambiguous signal of the diffusion wake in central Pb+Pb collisions. Projecting this two-dimensional distribution onto the rapidity direction within the azimuthal angle range $|\Delta\phi|<\pi/2$, one obtains a double-peak structure in the jet-hadron rapidity correlation. This provides a theoretical basis for experimental search for the diffusion wake signal by CMS and ATLAS. Both experiments have observed such structures in Z-jet~\cite{CMS:2025dua} and $\gamma$-jet~\cite{ATLAS:2024prm} events, respectively, providing the first direct experimental observation of jet-induced medium response in high-energy heavy-ion collisions.

$\gamma$ and Z-jets are relatively rare in current heavy-ion experiments at LHC with the limited beam luminosity, leading to large uncertainties in CMS and ATLAS measurements. It is therefore essential to explore the signal of jet-induced diffusion wake in dijet events, which are significantly more abundant at the LHC. The challenge in identifying the signal of the diffusion wakes induced by back-to-back dijets without a rapidity gap lies in the overlap of the diffusion wake induced by one jet with the soft hadron enhancement from the Mach-cone wake and the induced gluon radiation of the other jet in azimuthal angle. In this case, the diffusion wake primarily reduces the medium-induced hadron enhancement in the jet-hadron correlation of both jets, resulting in no clear signal of the diffusion wake. However, if one selects dijets with a finite rapidity gap, the diffusion wake valley of one jet will shift away from the Mach-cone wake of the other jet~\cite{Pablos:2019ngg}, leading to an asymmetrical jet-hadron rapidity correlation. In a recent study~\cite{Yang:2025dqu} , the 2-dimensional jet-hadron azimuthal angle and rapidity correlations were calculated in dijet events as in $\gamma$-jet events. The two back-to-back jets are required to have a finite rapidity gap $|\Delta\eta_{\rm jet_1,jet_2}|\equiv|\eta_{{\rm jet}_1}-\eta_{{\rm jet}_2}|>1$ as shown in the lower panel of Fig.~\ref{fig1}(c). The jet-hadron correlation in central Pb+Pb collisions exhibits two distinct peaks: one at $\Delta\phi = 0$ corresponding to the near-side correlation for hadrons from the leading (trigger) jet, and the other at $\Delta\phi = \pi$ corresponding to the away-side correlation for hadrons from the sub-leading jet. Since one requires a rapidity gap between the two jets, these two peaks are not located in the same rapidity region. Similar to $\gamma$-jets, these two peaks are enhanced relative to p+p collisions due to jet-induced medium response and medium-induced gluon radiation. Furthermore, the near-side jet-hadron rapidity correlation reveals a clear dip around $\Delta\eta$ = -2, which is absent in p+p collisions, due to the diffusion wake valley from the sub-leading jet.
To further reveal the effect of the diffusion wakes in dijets, one can project the correlations onto the rapidity in both near ($\Delta\phi<\pi/2$) and away-side ($\Delta\phi>\pi/2$) region of the leading jet, and focus on the difference or rapidity asymmetry between medium modifications of the projected jet-hadron correlations in dijets with rapidity gap $\Delta N_{\rm AA}$ and without rapidity gap $\Delta N_{\rm AA}(|{\rm\Delta\eta_{12}}|<0.5)$ as shown in Fig.~\ref{fig1}(d) for both near-side and away-side jet. One can see a depletion of soft hadrons in the shifted rapidity region of the diffusion wake and an enhancement in the rapidity region of the other jet whose soft hadron enhancement is no longer or less reduced by the diffusion wake as in dijets without a rapidity gap. This rapidity asymmetry is a direct consequence of the diffusion wakes induced by dijets. Note that contributions to jet-hadron correlations from the uncorrelated background of the bulk medium completely cancels in the rapidity asymmetry. This removes the need for background subtraction and the  associated uncertainties in direct measurement of the jet-hadron correlations.

In summary, 2-dimensional jet-hadron correlations in azimuthal and rapidity have been proposed recently to search for the signal of the jet-induced diffusion wake in high-energy heavy-ion collisions. In $\gamma$-jet events, a clear diffusion wake valley on top of an MPI ridge has been identified as the signal of jet-induced diffusion wake in theoretical simulations. This has led to the first direct experimental observation of the jet-induced medium response in central Pb+Pb collisions at LHC by CMS and ATLAS Collaborations. A unique rapidity asymmetry in the jet-hadron correlations that is background-free is also proposed as a robust signal of the diffusion wake induced by dijets with a finite rapidity gap. With large amount of dijet events at LHC and RHIC, more accurate measurements of the diffusion wakes through such asymmetry in rapidity become possible. Combined with the measurement of the diffusion wakes in Z/$\gamma$-jets, these studies are expected to advance our understanding of the jet-induced medium response and the properties of the QGP in high-energy heavy-ion collisions.

\noindent{\it\color{blue} Acknowledgment --} 
This work is supported in part by the China Postdoctoral Science Foundation under Grant No. 2024M751059 and No. BX20240134(ZY), by NSFC under Grant No. 12535010 and by the Guangdong MPBAR with No. 2020B0301030008. Computations in this study are performed at the NSC3/CCNU and NERSC under the award NP-ERCAP0032607.

\bibliography{ref}

\end{document}